\documentclass[conference,a4paper]{IEEEtran}
\IEEEoverridecommandlockouts
\usepackage[hidelinks]{hyperref}
\usepackage{cite}
\usepackage{amsmath,amssymb,amsfonts}
\usepackage{algorithmic}
\usepackage{graphicx}
\usepackage{textcomp}
\usepackage{xcolor}
\usepackage{amsfonts}
\usepackage{booktabs}
\usepackage{siunitx}
\usepackage{bbm}
\usepackage{multirow}
\usepackage[ruled, vlined]{algorithm2e}
\def\BibTeX{{\rm B\kern-.05em{\sc i\kern-.025em b}\kern-.08em
    T\kern-.1667em\lower.7ex\hbox{E}\kern-.125emX}}

\DeclareMathAlphabet{\mathcal}{OMS}{cmsy}{m}{n}
\SetMathAlphabet{\mathcal}{bold}{OMS}{cmsy}{b}{n}

\begin{document}

\title{Informed Graph Learning By Domain Knowledge Injection and Smooth Graph Signal Representation\\

}
\author{
    \IEEEauthorblockN{Keivan Faghih Niresi, Lucas Kuhn, Gaëtan Frusque, Olga Fink}
    \IEEEauthorblockA{Intelligent Maintenance and Operations Systems (IMOS) Lab, EPFL, Switzerland \\
 \
    \{keivan.faghihniresi, lucas.kuhn, gaetan.frusque, olga.fink\}@epfl.ch}
    
    \thanks{This research was funded by the Swiss Federal Institute of Metrology (METAS).}    
    }

\maketitle

\begin{abstract}
Graph signal processing represents an important advancement in the field of data analysis, extending conventional signal processing methodologies to complex networks and thereby facilitating the exploration of informative patterns and structures across various domains. However, acquiring the underlying graphs for specific applications remains a challenging task. While graph inference based on smooth graph signal representation has become one of the state-of-the-art methods, these approaches usually overlook the unique properties of networks, which are generally derived from domain-specific knowledge. Overlooking this information could make the approaches less interpretable and less effective overall. In this study, we propose a new graph inference method that leverages available domain knowledge. The proposed methodology is evaluated on the task of denoising and imputing missing sensor data, utilizing graph signal reconstruction techniques. The results demonstrate that incorporating domain knowledge into the graph inference process can improve graph signal reconstruction in district heating networks. Our code is available at \href{https://github.com/Keiv4n/IGL}{github.com/Keiv4n/IGL}.
\end{abstract}


\section{Introduction}
Spatial distribution of the collected data has emerged as an important property across a wide range of applications such as traffic data analysis\cite{bui2022spatial}, air pollution networks\cite{air2023}, and biological networks\cite{biological}. The intrinsic network characteristics of these datasets contain important insights into the connections and information between different entities across the network. Graphs are essential tools for representing the complex structures present in such data, as they provide flexible mathematical representations and can offer both  analytical and visual foundations  for understanding  and interpreting large amounts of data. In recent years, there has been an effort to extend signal processing techniques to graphs, resulting in the emergence of the graph signal processing (GSP) field, which aims to improve the data representation on graphs \cite{ortega2018graph, shuman2013graph}. However, acquiring the underlying graphs for specific  applications can be challenging, and constructing graphs based solely on network connectivity may not guarantee optimal results for certain subsequent tasks such as forecasting \cite{shang2021discrete}. Therefore, it is crucial to infer a graph that effectively captures the structure of the data. 

Graph inference is an ill-posed problem that aims to define the generative function  most accurately describing the relationship between the learned graph topology and the observed data \cite{matta2020graph}. Sample correlation, Gaussian radial basis function kernel, and cosine similarity are among the most straightforward methods for capturing the similarity of data samples \cite{qiao2018data}. However, these methods are vulnerable to noise because they rely solely on observations and do not utilize an explicit prior or data model. Therefore, different approaches have recently been proposed for graph learning based on GSP\cite{dong2019learning}. These techniques allow for the direct extraction and inference of underlying graph structures from the data. Graph inference methods based on GSP can be categorized into global smoothness-based \cite{dong2016learning, kalofolias2016learn}, dictionary-based \cite{thanou2017learning}, and spectral template-based methods \cite{segarra2017network}. In this study, we focus on global smoothness-based  methods for our proposed technique, ensuring its adaptability is preserved. This focus is motivated by the availability of scalable and efficient solvers within this category, which aligns well with the requirements of our proposed method which is mainly designed for large-scale sensor networks. Moreover, this type of method is explainable from both signal representation and statistical perspectives. For a comprehensive understanding of various  methods, interested readers are encouraged to explore relevant literature \cite{dong2019learning, mateos2019connecting}. 

Although global smoothness-based methods have demonstrated competitive results in graph inference, they neglect the unique characteristics of the physical processes in networks derived from domain knowledge, potentially limiting their overall effectiveness and interpretability. To address this limitation, this study proposes a novel graph inference method that leverages the presence of domain knowledge. To optimize the graph inference task, we efficiently solve the optimization problem using the primal-dual splitting algorithm. We evaluate the effectiveness of our approach on the task of graph signal reconstruction for denoising and imputing missing sensor data in the district heating network. The key contributions of our work include:

 % In \cite{dong2016learning}, a prior distribution intended to ensure signal smoothness on the graph is used in factor analysis to characterize the data. This approach  results in an optimization problem with constraints to ensure valid Laplacian matrix generation. In \cite{kalofolias2016learn}, the objective of the graph learning problem was reformulated to explore solutions within the space of a weighted adjacency matrix rather than a Laplacian matrix. Then, various regularization terms are incorporated to enhance the accuracy of graph inference, along with computationally efficient optimization techniques for improved scalability in different scenarios

 % We demonstrate how this domain knowledge can be integrated into existing graph learning methods based on smooth graph signal representation to capture the interaction among entities  in the absence of trivial descriptive relations from domain experts.

\begin{itemize}
    \item We propose a novel method for graph inference that leverages available domain knowledge.
    \item We present an efficient solution  of the optimization problem through the use of a primal-dual algorithm.
    \item We validate the effectiveness of our  proposed approach through  a case study on graph signal reconstruction, specifically in denoising and imputing missing data from district heating network sensors.
\end{itemize}

% \emph{Interdisciplinary Impact:} Traditional methods in civil engineering rely on nominal models for simulating and inferring unknown temperature and pressure values. However, the uncertainty in many physical parameters, such as pipe roughness and friction coefficient, makes the nominal model less readily available, necessitating  meticulous calibration. On the other hand, GSP offers an alternative approach, by reconstructing values based on graph signal smoothness. This approach, however,  depends on the fidelity of the graph topology and the accuracy of sensor values. Our work aims to bridge these two models, leveraging their strengths to compensate for each other's shortcomings.

\section{Preliminaries}
\label{sec:background}
\subsection{Graph Signal Processing}
We consider a weighted undirected graph $\mathcal{G} = (\mathcal{V}, \mathcal{E}, \mathbf{W})$, where $\mathcal{V}$, $\mathcal{E}$, and $\mathbf{W}$ represent the sets of nodes, edges, and the adjacency matrix, respectively. The graph's topology is defined  by the adjacency matrix $\mathbf{W}$ of size ${n \times n}$, with $\mathbf{W}{({i,j})}$ denoting  the edge weight between vertices $v_i$ and $v_j$. If  there is no edge between $v_i$ and $v_j$, the $\mathbf{W}{({i,j})}$ is set to zero.  The Laplacian matrix $\mathbf{L}$ is defined as $\mathbf{L}:= \mathbf{D} - \mathbf{W}$, where $\mathbf{D}$ is a diagonal matrix containing the degree of each node. 

Several studies have used the graph signal smoothness assumption to address  graph inference problems. The discrete \(p\)-Dirichlet form has been introduced as a notion of global smoothness in such studies as those by \cite{shuman2013graph, giraldo2022reconstruction}:

\begin{equation}
S_p(\mathbf{x}) = \frac{1}{p} \sum_{i \in \mathcal{V}} \|\nabla_i \mathbf{x}\|_p^2.
\label{dirichlet}
\end{equation}

For example, the well-known graph Laplacian quadratic form is achieved when \(p = 2\):

\begin{equation}
S_2(\mathbf{x}) = \sum_{(i, j) \in \mathcal{E}} \mathbf{W}(i, j) [\mathbf{x}(j) - \mathbf{x}(i)]^2 = \mathbf{x}^\mathsf{T}\mathbf{L}\mathbf{x}.
\label{vectorsmoothness}
\end{equation}

For multiple snapshots of graph signals, an extension of the graph Laplacian quadratic form can be expressed  as:

\begin{equation}
 S_2(\mathbf{X}) = \sum_{i=1}^{m} \mathbf{x}_i^\mathsf{T} \mathbf{L} \mathbf{x}_i = \mathsf{tr}(\mathbf{X}^\mathsf{T} \mathbf{L} \mathbf{X}), 
 \label{smoothness}
\end{equation}
where $\mathbf{x_i} \in \mathbb{R}^n$ represents  a graph signal at time $i$,  $m$ is the number of all available snapshots and $\mathsf{tr}(\cdot)$ denotes the trace operator.

\subsection{Graph Inference with Smooth Graph Signal Representation}
The main objective of learning the graph structure based on smooth graph signal representation is to minimize the Laplacian quadratic function (\ref{smoothness}). However, minimizing this function (\ref{smoothness}) with respect to the Laplacian matrix ($\mathbf{L}$) leads to the trivial solution of all edge weights being zero. To overcome this issue, regularization terms and constraints are introduced into the objective function \cite{dong2016learning} to estimate a valid $\mathbf{L}$:
\begin{equation}
\begin{aligned} 
&\min_{\mathbf{L}} && \mathsf{tr}(\mathbf{X}^\mathsf{T} \mathbf{L} \mathbf{X}) + \beta_1\|\mathbf{L}\|_{F}^{2} \\ 
&\text{s.t.} && \mathsf{tr}(\mathbf{L})=n, \\ 
&&& \mathbf{L}(i,j)=\mathbf{L}(j,i) \leq 0, \quad i \neq j, \\ 
&&& \mathbf{L} \mathbf{1}=\mathbf{0},
\end{aligned}
\label{eq:dong}
\end{equation}
where, $\beta_1$ is a regularization parameter, $\mathbf{1}$ denotes the constant vector of ones, and $\|\cdot\|_{F}$ represents the Frobenius norm of a matrix.

In \cite{kalofolias2016learn}, an alternative method was proposed for identifying a graph by exploring the space of weighted adjacency matrices, instead of  focusing solely on the Laplacian matrix. This approach leads to more straightforward and intuitive problem formulations, which can be solved more quickly and efficiently. %Additionally, they introduced a novel regularization technique that enforces positive degrees but does not restrict individual edges from becoming zero by the logarithmic barrier:
\begin{multline}
\min_{\mathbf{W}} \|\mathbf{W} \odot \mathbf{Z}\|_{1,1} - {\alpha_2}{\mathbf{1}}^\mathsf{T}\log(\mathbf{W1}) + {\frac{\beta_2}{2}} \|\mathbf{W}\|_F^2 \\
\text{s.t.}\quad \mathbf{W} \in \mathbb{R}_+^{n \times n}, \quad \mathbf{W}=\mathbf{W}^\mathsf{T},\quad \mathsf{diag}(\mathbf{W}) = \mathbf{0}.
\label{eq:kalo}
\end{multline}
The initial term represents the elementwise $\ell_1$ norm, aimed at promoting sparsity. Here, the weights are determined by the distance between elements within the signal, generating a distance matrix $\mathbf{Z}(i,j) = |\mathbf{x}(i) - \mathbf{x}(j)|^2$. Additionally, the symbol $\odot$ denotes the elementwise (Hadamard) product operation. The second term introduces a logarithmic barrier that enforces positive degrees but does not prevent  individual edges from becoming zero. $\alpha_2$ and $\beta_2$ are the regularization parameters, and the space of solutions is restricted by constraints to enforce a positive edge weight, undirected graph, and graph without self-loop. In \cite{kalofolias2016learn}, this optimization problem (\ref{eq:kalo}) has been solved efficiently by primal-dual algorithms. 
%Where the first term correspond to a weighted $\ell_1$ norm for sparsity, the weights correspond to the distance between the signal $i$ and $j$, which give the distance matrix $\mathbf{Z}(i,j) = \|\mathbf{x}(i) - \mathbf{x}(j)\|^2$ and $\odot$ is the elementwise (Hadamard) product. $\alpha_2$ and $\beta_2$ are the regularization parameters, and the space of solutions is restricted by constraints to enforce a positive edge weight, undirected graph, and graph without self-loop. In \cite{kalofolias2016learn}, this optimization problem (\ref{eq:kalo}) has been solved efficiently by primal-dual algorithms. 

\section{Proposed Methods}
\label{sec:method}
In this section, we first demonstrate how the characteristics of the physical processes in district heating networks can be interpreted as distances between nodes to construct a graph. Subsequently, we propose the informed graph learning (IGL) method by integrating this constructed graph into a smooth graph signal representation, enabling the learning of connections between nodes with limited domain knowledge. In summary, our approach learns a graph aligned with domain knowledge while leveraging smooth graph signal representation to uncover connections in less-explored areas.

\subsection{Physics-Inspired Graph Construction}
Learning the graph, instead of relying solely on the physical connectivity of networks, offers several  advantages. Firstly, physical connectivity graphs may not capture all relevant relationships and interactions among nodes, especially in complex systems where intricate dependencies exist. Moreover, they are less informative since they only  indicate the connectivity without assigning any weights to the edges. This issue is particularly evident in our case study, which focuses on a district heating network. Despite the presence of physical connectivity, it may not comprehensively capture the complex interactions between nodes. However, this specific network has been studied from a fluid dynamics perspective. Therefore, an alternative approach to graph learning, as opposed to relying solely on physical connectivity, involves constructing a network graph based on available domain knowledge, interpreting certain characteristics of the physical processes as connectivity strengths between two nodes.  District heating networks, typically equipped with pressure and temperature sensors for monitoring, can be represented as a graph by considering the variations in temperature and pressure drop along the pipes. Then, a stronger connection between the two sensors is established when there is a lower pressure and temperature drop along the pipe between them. We calculate pressure drop ($\left|\Delta P_{ij}\right|$) along two pressure nodes $v_i$ and $v_j$ by the Hazen-Williams equation as:

\begin{equation}
\left|\Delta P_{ij}\right| = \left|\frac{10.67 \cdot L_{ij} \cdot Q_{ij}^{1.852}}{R_{ij}^{1.852} \cdot D_{ij}^{4.87}}\right|,
\end{equation}
where $L_{ij}$, $D_{ij}$, $R_{ij}$, $Q_{ij}$ represents the pipe length, diameter, Hazen-Williams roughness coefficient, and volumetric flow of the pipe which connects node $v_i$ to $v_j$. The temperature drop ($\left|\Delta T_{ij}\right|$) between node $v_i$ and $v_j$ can be approximated as:

\begin{equation}
\left|\Delta T_{ij}\right| \approx \left|\frac{\dot{q}_{ij}}{\dot{m}_{ij} \cdot C_{ij}}\right|,
\end{equation}
where $\dot{q}_{ij}$, $C_{ij}$, $\dot{m}_{ij}$ represent the heat transfer rate, the specific heat capacity of water, and mass flow rate, respectively.

After calculation of $\left|\Delta P_{ij}\right|$ and $\left|\Delta T_{ij}\right|$, the graph can be constructed separately for each of the pressure ($\mathbf{W}^{p}(i,j)$) and temperature sensors ($\mathbf{W}^{t}(i,j)$) such that:
\[
\mathbf{W}^{p}(i,j) = \frac{1}{\left|\Delta P_{ij}\right|}, \quad \mathbf{W}^{t}(i,j) = \frac{1}{\left|\Delta T_{ij}\right|}.
\]

Since the values of temperature and pressure drops are in different ranges, we rescale $\mathbf{W}^{p}(i,j)$ and $\mathbf{W}^{t}(i,j)$ separately, such that the edge weights are between 0 and 1. Then, we eliminate edges with weights below 0.1 to enforce sparsity. Finally, to merge the scaled pressure graph (\(\mathbf{W}^p_s(i,j)\)) and temperature graph (\(\mathbf{W}^t_s(i,j)\)) into the overall physics-inspired graph (\(\mathbf{W_{PI}}\)), we combine them into a block-diagonal matrix:
\begin{equation}
\mathbf{W_{PI}} = \begin{bmatrix}
    \mathbf{W}^p_s & \mathbf{0} \\
    \mathbf{0} & \mathbf{W}^t_s
\end{bmatrix}.
\label{W_PI}
\end{equation}

\begin{figure}
    \centering
    \includegraphics[width=0.45\linewidth]{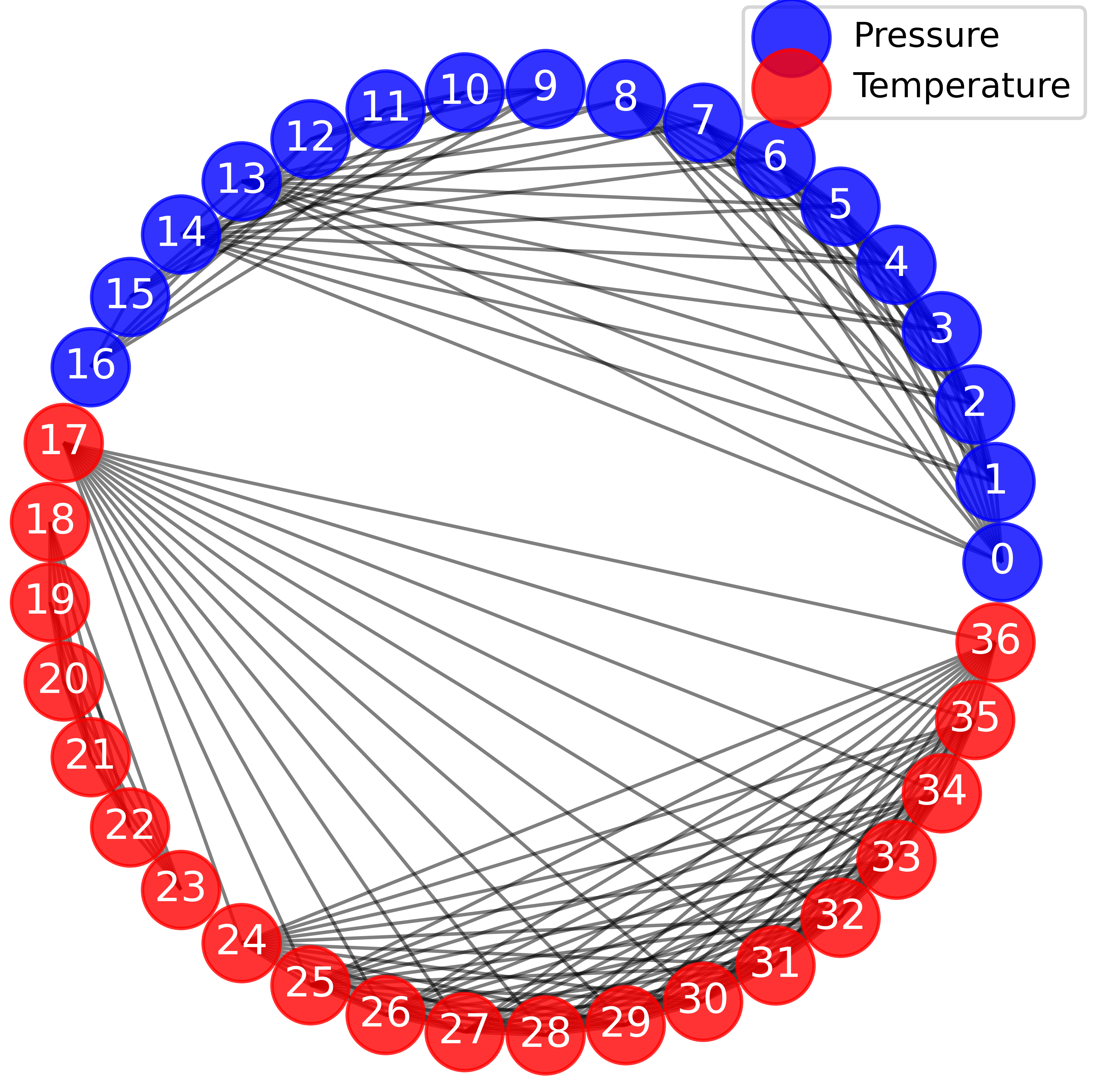}
    \caption{Constructed graph based on characteristics of the physical processes.}
    \label{fig:physicsgraph}
    \vspace{-2mm}
\end{figure}

Equation (\ref{W_PI}) describes a unified graph that incorporates two distinct subgraphs, corresponding to pressure and temperature sensors, essentially forming two disconnected graphs within the larger structure. For better intuition, Figure \ref{fig:physicsgraph} illustrates the graph constructed based on characteristics of the physical processes. It is evident that, due to the absence of domain knowledge connecting pressure to temperature nodes (or vice versa), the entire graph includes distinct subgraphs for each sensor type.

\subsection{Proposed Formulation for Informed Graph Learning}
Once $\mathbf{W_{PI}}$ is obtained based on domain knowledge, we can address  a new optimization problem by incorporating additional regularization into Equation (\ref{eq:kalo}) to obtain  the following:
\vspace{-2mm}
\begin{multline}
\min_{\mathbf{W}} \|\mathbf{W} \odot \mathbf{Z}\|_{1,1} - {\alpha}{\mathbf{1}}^\mathsf{T}\log(\mathbf{W1}) \\
+ {\frac{\beta}{2}} \|\mathbf{W}\|_F^2 + {\frac{\upsilon}{2}} \| \mathbf{M} \odot \mathbf{W} - \mathbf{W_{PI}}\|_F^2 \\
\text{s.t.}\quad \mathbf{W} \in \mathbb{R}_+^{n \times n}, \quad \mathbf{W}=\mathbf{W}^\mathsf{T},\quad \mathsf{diag}(\mathbf{W}) = \mathbf{0},
\label{eq: Proposed}
\end{multline}
where $\mathbf{M}$ is the physical knowledge index matrix, indicating the links for which we have domain knowledge, such that: 
\begin{equation}
    \mathbf{M}(i,j) = \begin{cases}
  1  & \mathbf{W_{PI}}(i,j) \neq 0\\
  0 & \mathbf{W_{PI}}(i,j) = 0
\end{cases}.
\end{equation}
Equation (\ref{eq: Proposed}) specifies that the graph is learned with consideration for domain knowledge, as indicated by $\mathbf{M}$. The objective is to ensure consistency in the parts where domain knowledge is available ($\mathbf{M}(i,j) = 1$). For the parts without domain knowledge ($\mathbf{M}(i,j) = 0$), the goal is to rewire the graph using smooth graph signal representation. In summary, domain knowledge and smooth graph signal representation complement each other in the construction of a new graph,  resulting in a graph signal that is both smooth and consistent with the provided domain knowledge.

\subsection{Optimization}
The optimization problem specified in Equation (\ref{eq: Proposed}) can be efficiently solved using various algorithms. Before deriving the update steps for this problem, it is essential to note that due to the symmetry of the matrix $\mathbf{W}$ (second constraint) and the absence of self-loops (third constraint), the problem can be effectively solved by focusing solely on the upper triangular part of (\(\mathbf{W}(i,j), j > i\)). This implies that instead of addressing the problem in $\mathbb{R}_{+}^{n \times n}$, it can be tackled in $ \mathbf{w} \in \mathbb{R}_{+}^{n(n-1)/2}$ without explicit consideration of the second and third constraints. Additionally, similar to \cite{kalofolias2016learn}, we incorporate an indicator function (\(\mathbbm{1}\{\mathbf{w} \succeq \mathbf{0}\} = 0\) if \(\mathbf{w} \succeq \mathbf{0}\), and \(\mathbbm{1}\{\mathbf{w} \succeq \mathbf{0}\} = \infty\) otherwise) as a penalty function to enforce non-negativity constraints. With these definitions, we reformulate the objective in equation (\ref{eq: Proposed}) as:

\begin{multline}
\min_{\mathbf{w}} \left( \mathbbm{1}\{\mathbf{w} \succeq \mathbf{0}\} + 2\mathbf{w}^\mathsf{T}\mathbf{z} - \alpha \mathbf{1}^\mathsf{T}\log(\mathbf{d}) \right. \\
+ \left. \beta \|\mathbf{w}\|^2 + \upsilon \| \mathbf{m} \odot \mathbf{w} - \mathbf{w_{PI}}\|^2 \right),
\label{eq: vecProposed}
\end{multline}
where $\mathbf{d} \in \mathbb{R}_{+}^n$ represents  the vector of node degrees. To adapt the optimization problem (\ref{eq: vecProposed}) for primal-dual algorithms \cite{komodakis2015playing}, we divide  the objective into  the sum of three functions to utilize  the Monotone+Lipschitz Forward Backward Forward (M+LFBF) algorithm:

\begin{equation}
\min_{\mathbf{w}} f(\mathbf{w}) + g(\mathbf{Sw}) + h(\mathbf{w}),
\label{primaldual}
\end{equation}
where \(h\) is required to be differentiable with a gradient that possesses  a Lipschitz constant $\zeta$. The functions \(f\) and \(g\) should be such that their proximal operators are  readily accessible. Owing to \(\mathbf{S}\) being a  linear operator, \(g\) is defined on the dual variable ($\mathbf{Sw = d = W1} \in \mathbb{R}^n$). Finally, based on (\ref{eq: vecProposed}) and (\ref{primaldual}), we can delineate  and define $f$, $g$, and $h$ in the following way:

\begin{equation}
f(\mathbf{w}) = \mathbbm{1}\{\mathbf{w} \succeq \mathbf{0}\} + 2\mathbf{w}^\mathsf{T}\mathbf{z},
\end{equation}
\begin{equation}
g(\mathbf{d}) = - \alpha \mathbf{1}^\mathsf{T}\log(\mathbf{d}),
\end{equation}
\begin{equation}
h(\mathbf{w}) = \beta \|\mathbf{w}\|^2 + \upsilon \| \mathbf{m} \odot \mathbf{w} - \mathbf{w_{PI}}\|^2 \quad \zeta = 2(\beta + \upsilon).
\end{equation}

Finally, to derive the optimization step, we have:

\begin{equation}
\mathbf{prox}_{\lambda f}(\mathbf{y}) = \max(\mathbf{0}, \mathbf{y} - \lambda \mathbf{z}), \quad \text{elementwise}
\end{equation}
\begin{equation}
\mathbf{prox}_{\lambda g}(\mathbf{y}) = \frac{\mathbf{y} + \sqrt{\mathbf{y}^2 + 4\alpha\lambda}}{2}, \quad \text{elementwise}
\end{equation}
\begin{equation}
\nabla h(\mathbf{w}) = 2\beta \mathbf{w} + 2 \upsilon(\mathbf{m}\odot \mathbf{w} - \mathbf{w_{PI}}).
\end{equation}

Algorithm 1 provides a comprehensive summary of the informed graph learning (IGL) method.

\begin{algorithm}
  \SetKwInOut{Input}{Input}
  \SetKwInOut{Output}{Output}
  \SetKwInOut{Init}{Initialize}
    \caption{M+LFBF Algorithm for IGL (\ref{eq: vecProposed})}
    \Input{$\alpha, \beta, \gamma \in (0, 1 + \zeta + \|\mathbf{S}\|), \upsilon ,\epsilon_0, \mathbf{z}, \mathbf{S}, \mathbf{m}, \mathbf{w_{PI}}$}
    \Init{$\mathbf{w}^0 \in \mathbb{R}_+^{n(n-1)/2}, \mathbf{d}^0 \in \mathbb{R}_+^{n}$}
    \For{$k = 1, \ldots, k_{\max}$}{
        $\mathbf{y}^k = \mathbf{w}^k - \gamma(2\beta \mathbf{w}^k + 2 \upsilon(\mathbf{m}\odot \mathbf{w}^k - \mathbf{w_{PI}}) + \mathbf{S}^\mathsf{T}\mathbf{d}^k)$\;
        $\bar{\mathbf{y}}^k = \mathbf{d}^k + \gamma(\mathbf{S}\mathbf{w}^k)$\;
        $\mathbf{p}^k = \max(\mathbf{0}, \mathbf{y}^k - 2\gamma \mathbf{z})$ \quad \text{\# elementwise}\;
        $\bar{\mathbf{p}}^k = (\bar{\mathbf{y}}^k - \sqrt{(\bar{\mathbf{y}}^k)^2 + 4\alpha\gamma})/2$ \quad \text{\# elementwise}\;
        $\mathbf{q}^k = \mathbf{p}^k - \gamma(2\beta \mathbf{p}^k + 2 \upsilon(\mathbf{m}\odot \mathbf{p}^k - \mathbf{w_{PI}}) + \mathbf{S}^\mathsf{T}\mathbf{p}^k)$\;
        $\bar{\mathbf{q}}^k = \bar{\mathbf{p}}^k + \gamma(\mathbf{S}\mathbf{p}^k)$\;
        $\mathbf{w}^k = \mathbf{w}^k - \mathbf{y}^k + \mathbf{p}^k$\;
        $\mathbf{d}^k = \mathbf{d}^k - \bar{\mathbf{y}}^k + \bar{\mathbf{q}}^k$\;
        \If{$\frac{\|\mathbf{w}^k - \mathbf{w}^{k-1}\|}{\|\mathbf{w}^{k-1}\|} < \epsilon_0$ and $\frac{\|\mathbf{d}^k - \mathbf{d}^{k-1}\|}{\|\mathbf{d}^{k-1}\|} < \epsilon_0$}{
            \textbf{break}\;
        }
    }
    \Output{$\mathbf{w}^k$}
\end{algorithm}
After solving the problem for the upper triangular part of the weighted adjacency matrix through vectorization, we can reconstruct the symmetric adjacency matrix $\mathbf{W}$. The complexity of Algorithm 1 is $\mathcal{O}(n^2)$ for each iteration with $n$ nodes, and it  can be executed  in parallel.

\section{Experimental Results}
\label{sec:results}

Due to the absence of publicly available real-world datasets for district heating networks, we have created  a synthetic dataset consisting of 8760 samples, using the TesPy library \cite{witte2020tespy} for this purpose. The first 5000 samples are allocated for training, while the remaining 3760 samples are used for testing.  Min-max normalization is applied separately to pressure and temperature sensors, based on the minimum and maximum values of the training set. To enhance the realism of the synthetic dataset, zero-mean Gaussian noise with a standard deviation ($\sigma$) of 0.25 is added to the training data.

For hyperparameter tuning, 5-fold cross-validation is employed on the training data to select the optimal parameters based on denoising performance\footnote[1]{$\beta = 0.4$ captures edge density patterns, and $\upsilon = 0.4$ signifies the fidelity of the learned graph to domain knowledge. The optimization stopping criterion, $\epsilon_0$, is set to $10^{-5}$. Moreover, the learned adjacency matrix is normalized through elementwise division of each entry by the maximum edge value, followed by thresholding to drop weak edges with values less than $0.1$.}. 

\begin{table}
    \centering
    \caption{Quantitative Comparison of Methods for Graph Signal Reconstruction}
    \label{Imputation}
    \resizebox{\linewidth}{!}{
        \begin{tabular}{ccccccc} 
            \toprule
            Scenario & Metric  & Physics & Lap-Smooth & Adj-Smooth & IGL \\
            \toprule

            \multirow{2}{*}{Denoising ($\sigma = 0.3$)}
            & RMSE  &  2.829 & 2.421 & 2.403 & \textbf{2.395} &  \\ 
            & MAE  & 1.716 & 1.475 &  1.466 & \textbf{1.460} & \\ 

            \midrule

            \multirow{2}{*}{Imputation ($\rho = 0.3$)} 
            & RMSE  &  5.761 & 2.164 & 1.824 & \textbf{1.813} &  \\ 
            & MAE  & 1.888 & 1.179 &  0.987 & \textbf{0.977} & \\ 

            \midrule
            \multirow{2}{*}{Imputation ($\rho = 0.5$)} 
            & RMSE  & 2.558 & 1.846 & 1.489 & \textbf{1.466} & \\ 
            & MAE  & 1.056 & 1.091 & 0.885 & \textbf{0.869} &  \\ 

            \midrule
            \multirow{2}{*}{Imputation ($\rho = 0.7$)} 
            & RMSE  & 1.567 & 1.774 & 1.425 & \textbf{1.393}  \\ 
            & MAE  & 0.911 & 1.055 & 0.846 & \textbf{0.824}  \\ 

            \midrule
            \multirow{2}{*}{Imputation ($\rho = 0.9$)} 
            & RMSE  & 1.516 & 1.746 & 1.400 & \textbf{1.359}  \\ 
            & MAE  & 0.909 & 1.049 & 0.836 & \textbf{0.808} \\ 

            \bottomrule
        \end{tabular}
    }
    \vspace{-2mm}
\end{table}

\begin{figure}
    \centering
    \includegraphics[width=0.8\linewidth]{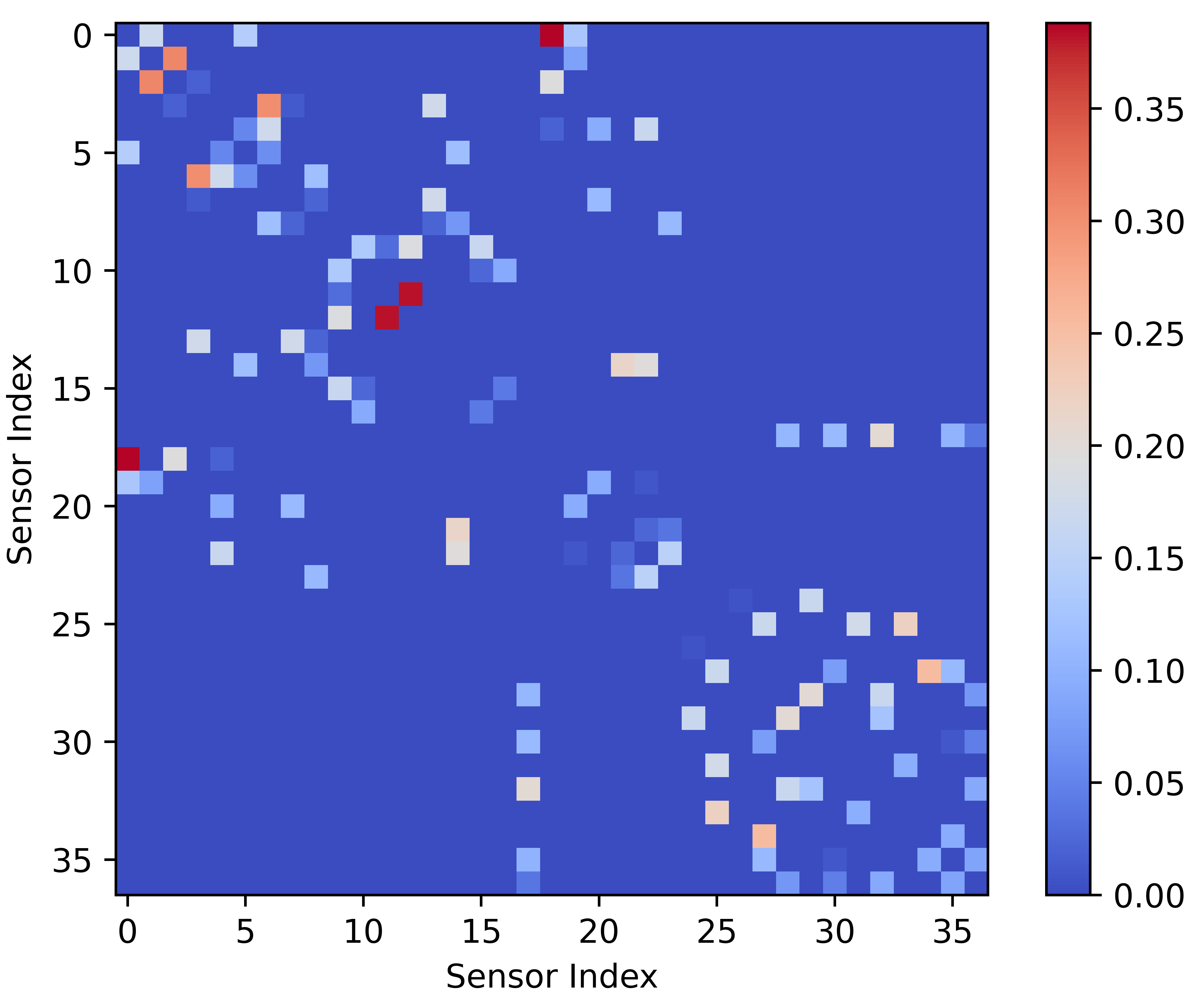}
    \caption{Colormap showing the absolute difference between IGL and Adj-Smooth adjacency matrices. Nodes 0-16 are pressure sensors, and nodes 17-36 are temperature sensors.}
    \vspace{-2mm}
    \label{fig:heatmap}
\end{figure}

For comparison, we evaluate our proposed IGL algorithm against a pure domain knowledge approach (\ref{W_PI}) based on characteristics of the physical processes, referred to as `Physics' in the results, smoothness optimization on the Laplacian matrix, referred to as Lap-Smooth\cite{dong2016learning}, and smoothness optimization on the adjacency matrix, referred to as Adj-Smooth \cite{kalofolias2016learn}. For evaluation, the learned graph from each method is utilized in denoising and imputation tasks by solving the convex optimization problem in the Appendix. For denoising, zero-mean Gaussian noise with a standard deviation of 0.3 is added to the test data. Additionally, four different cases related to various sampling densities ($\rho$) are considered for missing data imputation, where $\rho$ represents the fraction of available sensor data measurements.  The evaluation metrics for both tasks include root-mean-square error (RMSE) and mean absolute error (MAE). Table \ref{Imputation} presents the results for both imputation and denoising tasks. It can be observed that the graph constructed based on the underlying physics exhibits a significant performance drop as the number of missing values increases (for $\rho = 0.3$ and $0.5$). However, its performance remains competitive compared to other methods as the sampling ratio increases, attributed to the graph's limitation of considering only similar sensor types. This limitation prevents it from capturing complex interactions among different sensor types, thus hindering its effectiveness in scenarios  of low sampling densities. Moreover, the performance of Adj-Smooth and the proposed IGL method shows strong competitiveness. However, at higher sampling ratios ($\rho = 0.7$ and $0.9$) for the imputation task, the performance gap widens, with the proposed IGL method outperforming Adj-Smooth. This advantage comes from the additional regularization proposed, inspired by the characteristics of physical processes in district heating networks.
For  a more comprehensive comparison, the absolute difference between adjacency matrices of IGL and Adj-Smooth is visually represented by colormap in Fig. \ref{fig:heatmap}.

% \begin{table}
%     \begin{center}
%     \caption{Comparison of SNR, RMSE and MAE Metrics for Denoising Task.}
%     \label{tab:denoising}
% \begin{tabular}{cccc} \toprule
%     {Methods} & {SNR} & {RMSE} & {MAE} \\ \midrule
%     Noisy  & 16.133 & 7.887 &  4.832  \\
%     Physics  & 25.041 & 2.829 & 1.716  \\
%     Lap-Smooth  & 26.394 & 2.421  & 1.475   \\
%     Adj-Smooth  & 26.457 & 2.403  & 1.466  \\
%     IGL & \textbf{26.485}  & \textbf{2.395} & \textbf{1.460}\\ 
%                                                  \bottomrule
% \end{tabular}
% \end{center}
% \end{table}

% \begin{figure}
% \centering
% \begin{minipage}[b]{.4\textwidth}
% \centering
%  \includegraphics[width=.6\linewidth]{Kalo_Graph.png}
% \caption{Learned Graph by Adj-Smooth}\label{Kalo}
% \end{minipage}\qquad
% \begin{minipage}[b]{.4\textwidth}
% \centering
% \includegraphics[width=.6\linewidth]{Proposed_Graph.png}
% \caption{Learned Graph by IGL}\label{Proposed}
% \end{minipage}
% \end{figure}

% \begin{figure}

% \centering
% \begin{minipage}[b]{0.4\linewidth}
%   \centering
%   \includegraphics[width=\linewidth]{Kalo_Graph.png}
%   (a)
% \end{minipage}\qquad
% \begin{minipage}[b]{.4\linewidth}
%   \centering
%   \includegraphics[width=\linewidth]{Proposed_Graph.png}
%   (b)
% \end{minipage}
% \caption{(a): Graph learned by the Adj-Smooth, (b): Graph learned by the IGL}
% \label{fig: Graphs}
% \end{figure}

\section{Conclusion}
This study proposes a novel method for informed graph learning by using the characteristics of the physical processes in district heating networks and smooth graph signal representation.  The efficacy of the proposed approach was demonstrated through graph signal reconstruction, resulting in performance improvement relative to the compared methods. To the best of our knowledge, this work represents the first exploration of GSP in district heating networks. The proposed method can also be applied to other networks, such as power grids, where the graph can be constructed based on voltage drops among nodes. Future research directions may include assessing the proposed method in other tasks. In summary, the insights presented in this paper contribute to the progress of interdisciplinary research in signal processing.
\label{sec:conclusion}

\appendix
\subsection{Graph Signal Reconstruction}
Following graph inference, various inverse problems for graph signals can be addressed. Our emphasis in this study lies in the denoising and imputation. Specifically, for denoising purposes, the optimization problem takes the form:
\begin{equation}
\min_{\mathbf{\mathbf{X}}}  \|\mathbf{Y}-\mathbf{X} \|_F^2 + \mu\mathsf{tr}(\mathbf{X}^\mathsf{T} \mathbf{L} \mathbf{X}),
\label{eq:denoising}
\end{equation}
where $\mathbf{Y}$ is noisy data observation. Remarkably, a closed-form solution exists for this optimization problem, which is:
\begin{equation}
\mathbf{X} = (\mathbf{I} + \mu \mathbf{L})^{-1}\mathbf{Y}.
\end{equation}
Since the matrix $\mathbf{I} + \mu \mathbf{L}$ is positive definite, the inverse of this matrix can be efficiently computed through Cholesky decomposition \cite{dong2016learning, boyd2004convex}. 

For graph signal imputation, one can solve the following optimization problem:
\begin{equation}
\min_{\mathbf{\mathbf{X}}} \quad \frac{1}{2}\mathsf{tr}(\mathbf{X}^\mathsf{T} \mathbf{L} \mathbf{X}) \quad \text{s.t.}  \quad \mathbf{J}\odot\mathbf{X} = \mathbf{Y} ,
\label{eq:imputation}
\end{equation}
where $\mathbf{Y}$ is our observation with some missing values caused by sampling matrix $\mathbf{J}$. The solution to this problem (\ref{eq:imputation}) can be achieved through the gradient projection algorithm with the following iterative update:
\begin{equation}
\mathbf{X}^{k+1} = \mathcal{P}_\mathbf{Y}\left(\mathbf{X}^k - \xi \nabla_{\mathbf{X}} f_n(\mathbf{X}^k)\right),
\end{equation}
where \(f_n(\mathbf{X}^k) = \frac{1}{2} \mathsf{tr}\left((\mathbf{X}^k)^\mathsf{T}\mathbf{L} \mathbf{X}^k\right)\), \(\xi\) is the step size, \(\nabla_{\mathbf{X}} f_n(\mathbf{X}^k)\) is the gradient of the function \(f_n(\mathbf{X}^k)\) given by
\begin{equation}
\nabla_{\mathbf{X}} f_n(\mathbf{X}^k) = \mathbf{L}\mathbf{X}^k,
\end{equation}
and $\mathcal{P}_\mathbf{Y}(\mathbf{A})$ is the projection of $\mathbf{A}$ to space $\mathbf{Y}$   given by $\mathcal{P}_\mathbf{Y}(\mathbf{A}) = \mathbf{Y} + \mathbf{A} - \mathbf{J} \odot \mathbf{A}$.

% \section*{Acknowledgment}
% This research was funded by the Swiss Federal Institute of Metrology (METAS).
% \pagebreak

\bibliographystyle{IEEEtran}
\bibliography{IEEEabrv,References}

\end{document}